# GROWTH PATTERNS of MICROSCOPIC BRAIN TUMORS

Leonard M. Sander[1,] and Thomas S. Deisboeck[2,3]

[1] Department of Physics and Michigan Center for Theoretical Physics, The University of Michigan, Ann Arbor, MI, 48109.

[2] Complex Biosystems Modeling Laboratory, Harvard-MIT (HST) Athinoula A. Martinos Center for Biomedical Imaging, HST-Biomedical Engineering Center, Massachusetts Institute of Technology, Cambridge, MA 02139.

[3] Molecular Neuro-Oncology Laboratory, Massachusetts General Hospital, Harvard Medical School, Charlestown, MA 02129.

## ABSTRACT

Highly malignant brain tumors such as Glioblastoma Multiforme (GBM) form complex growth patterns *in vitro* in which invasive cells organize in tenuous branches. Here, we formulate a chemotaxi*s* model for this sort of growth. A key element controlling the pattern is homotype attraction, i.e., the tendency for invasive cells to follow pathways previously explored. We investigate this in two ways: we show that there is an intrinsic instability in the model, which leads to branch formation. We also give a discrete description for the expansion of the invasive zone, and a continuum model for the nutrient supply. The results indicate that both strong heterotype chemotaxis *and* strong homotype chemo-attraction are required for branch formation within the invasive zone. Our model thus can give a way to assess the importance of the various processes, and a way to explore and analyze transitions between different growth regimes.

## 1. Introduction

In certain three-dimensional tissue culture settings, highly malignant human brain tumors can form a remarkable growth pattern, consisting of central proliferation and invasive spread into the periphery[1]. In the brain itself, these tumors may grow in a similar way. In the initial stages the tumor grows more or less symmetrically until (presumably) mechanical confinement pressure increases to a critical level, and an invasive phase is triggered. In this phase the central multicellular tumor spheroid (MTS) continues to grow, but also there is rapid invasion of surrounding tissue by mobile tumor cells, which are continually shed by the MTS at several stages. In these stages chains of single cells can branch and extend around the core in an invasive zone. It is this rapid and extensive invasion of brain parenchyma surrounding the main macroscopic tumor, which makes this terrible disease so difficult to treat [2]. This paper will introduce a model that has some of the salient features of this process.

Some of the features of the growth and invasion have been revealed by *in vitro* experiments in transparent extracellular matrix gel[1]: see Figure 1. The branching pattern surrounding the central core is the invasive zone, which grows with a maximum velocity of about 4.5 µm/hr, which in turn corresponds well with values obtained from *in vivo* experiments[3]. Our interest here is to try to understand the mechanism for the formation of the branching structures.

It seems clear that we are dealing here with a growth instability. In fact, there is considerable qualitativ*e* resemblance between the pattern of Figure 1 and the well-known diffusion-limited aggregation (DLA) pattern[4] or biological patterns such as those encountered in nutrient-limited growth of bacteria colonies[5]. Modeling of a similar kind has been used for cancers other than brain tumors[6]. In these biological cases the basic pattern is determined by enhanced proliferation of tumor cells at well-nourished tips of branches. This leads to unstable growth of the branches exactly as in the classic Mullins-Sekerka instability of metallurgy[7].

However, the biology of brain tumors is different from that of bacteria colonies in the important respect that the invasive cells are thought to exhibit very little proliferative activity [8,9]. This means, that all (or most) of the new cells are produced at the surface of the MTS, and then invade the surrounding tissue. Secondary tumors may form at distant sites (compare also with Deisboeck *et al.,* [1]), yet the branch formation in Figure 1 is supplied from the MTS core, or more accurately from its highly proliferative cell surface layers (see also next section). The invasive cells move in response to various chemical and material gradients in the tissue and, we assume that there is a similar mechanism for the *in vitro* assay. That is, their motion is governed by chemotaxis and haptotaxis, though, in what follows we will neglect haptotaxis.

In our model, we will assume that the growth of the invasive branches is governed by two major processes: (i) *Chemotaxis* caused by the gradient of (heterotype) nutrient concentration. In the experiment, a (non-replenished) nutrient medium is mixed with the bio-gel, and it is consumed by the growing tumor (see the next section for details). This, however, would merely lead to an expanding cloud of mobile cells. (ii) *Homotype attraction*, another form of chemotaxis where cells secrete a soluble agent (paracrine production) which attracts other cells. There also may be tissue damage by the invading cells, which gives rise to pathways which other cells can follow more easily.

In the next section we give more details about the biological background of experiments and concepts. We show, for a simplified continuum model that we can get branch formation from a combination of heterotype and homotype attraction (both are necessary). We then introduce a hybrid discrete-continuum model for the tumor expansion. Our simulation results, in the following section, show that there are regimes of branch formation depending on the strength of various effects, and these are, in some measure, found in the experiment. Finally we summarize our results, and point out directions for future modeling efforts.

## 2. Experimental Background

### 2.1. Experimental Results

We developed an experimental model, which uses multicellular tumor spheroids (MTS [10]) implanted into a 3D extracellular matrix (ECM) gel. This sandwich MTS assay is described in detail in Deisboeck *et al.*[1] In brief, we placed human U87MG EGFR[11] multicellular glioma spheroids ( = 500-700 µm; 0.7-1.0 x $10^4$ cells) in between two layers of growth factor reduced (GFR) matrix, Matrigel (BIOCAT , Becton

Dickinson, Franklin Lakes, NJ), which forms a reconstituted basement membrane at room temperature. It has been shown that such basement membranes have distinct network structure[12]. This specific GFR-matrix variant contains less growth factor (e.g., EGF, PDGF, TGF- ) as compared to the full Matrigel, however a similar amount of the ECM proteins laminin (61%), collagen IV (30%) and entactin (7%). We then mixed this gel with (serum-free) OPTI-MEM to a ratio of 3:1 GFR-M to medium, reaching a total volume of 200 µl (per well; using a 48-well flat bottom tissue culture treated Multiwell plate (FALCON , Fisher Scientific, Pittsburgh, PA)).

As reported previously in Deisboeck et al.[1], over 144 hrs, the volumetric growth followed decelerating kinetics reaching on average 0.403 mm$^3$. During the same period invasion increased significantly up to 1.705 mm$^2$ at the end of the observation period. The experimental model also showed a steep increase in invasive edge cell velocity reaching a peak of 109 µm/day at t = 96 hrs. Specific immunohistochemistry staining revealed an inverse relationship between MTS size and proliferative index. Also, the proliferative cells tended to be more densely arranged in the surface layers of the MTS whereas in the center of the larger MTS cells were less dense with separate lucid areas and apoptotic nuclei.

Figure 1 depicts two microscopic sample images from tumor spheroids at 120 hours (after placement into the 3D ECM gel (using the same gel composition in both experiments)). Figure 1a shows "chain-like" invasion patterns, whereas the MTS system in Figure 1b displays a more "disc-like" invasive pattern. Invasion in both MTS experiments is predominantly 2D because of the design of the experiment[1]. At this time point, the average MTS volume was 0.356 mm$^3$ and the average MTS invasion area was 1.394 mm$^2$.

## 2.2. Modeling Background

Using the methods of the previous section we were able to study the evolving tumor with its key features of proliferation and invasion for up two six to eight days without replenishing the nutrient supply. Volumetric growth follows Gompertzian dynamics[13] Tumor growth and invasion were closely related and in fact showed evidence of feedback[1], (which we do not treat in this paper.) We can see in Figure 1 that invasive tumor cells often seemed to follow one another forming a chain, previously only shown for neural precursor cells[14]. We take this as evidence for homotype attraction of cells for one another. The biological correlate for the attractant(s) is the paracrine production of soluble protein growth factors, e.g., TGF- and HGF/SF by tumor cells [15,16,17]. Mobile cells would follow each other because of increasing paracrine attraction. There is also the possibility of smaller mechanical resistance in a preformed channel. Such a mechanism would lead to a continuous imprinti*ng* of (initially random) invasive pathways.

## 3. Formulation of the Model

### 3.1. Continuum Modeling

The major features that we want to include in our description of the mobile tumor cells are the following: cells are shed from the MTS and, in the absence of other forces, undergo *random motion*, i.e. diffusion. In continuum terms, the cells obey a diffusion equation with a diffusion coefficient, $D_c$. In addition, there is *chemotaxis*, i.e. directed active motion along chemical gradients. These effects are described by the usual Keller-Segel equation [18]:

$$\partial c/\partial t = \nabla \cdot [D_c(\mathbf{r}) \nabla c] - \nabla \cdot [c \, \chi(n,\mathbf{r}) \nabla n] - \nabla \cdot [c \, \mu \nabla h] \qquad (3.1)$$

Here $c$ is the concentration of mobile cells, and $n$, the concentration of nutrient which guides the motion of the cells. For the chemotaxis coefficient, $\chi$, we use the receptor law: $\chi = \chi(\mathbf{r}) n_o^2/(n_o + n)^2$. The characteristic concentration, $n_o$, will be discussed further, below. Here, $\chi$ gives the scale of the chemotaxis, i.e., the *drift velocity*, which a cell acquires in unit gradient of nutrient. This drift velocity accounts for a roughly linear growth law of the invasive zone. The last term is the homotype attraction mediated by a factor whose concentration is $h$, and whose chemotaxis coefficient (taken to be constant for simplicity) is $\mu$.

In our discrete simulation, below, we will allow both $D_c$ and $\chi$ to depend on position. This is to take into account the effect of tissue damage: we simply make it easier for a cell to move where other cells have previously been by multiplying $\chi$ and $D_c$ by a dimensionless factor, $\alpha_1$, which will be a parameter. This is a very crude representation of these effects, but it will allow us to show, in a fairly simple way, how branch formation occurs in a discrete model.

The boundary condition on *c* at the surface of the MTS corresponds to generation of cells as they are shed. As described above, we do not consider proliferation of cells in the invasive zone, based on experimental evidence [8,9]. In this respect our approach differs sharply from that of Burgess *et al.* [19] who, instead, write an equation of the form:

$$\partial c/\partial t = \nabla \cdot [D_c \nabla c] + gc(c_o - c) \qquad (3.2)$$

with logistic *growth throughout the invasive zone*, and *no* chemotaxis. This equation is of the form of a Fisher-Kolmogorov[20] equation, and has traveling front solutions with a fixed velocity,

$$v = 2\sqrt{D_c g c_o} \qquad (3.3)$$

due to the exponential growth of *c*. The authors use this equation to compute the diffusion coefficient.

However, our point of view is quite different. If we adopt an equation of the form of (3.2), we would have to put the growth term only on the surface of the MTS. We can see the difficulty with this if we idealize the MTS as a point source of new cells, and write:

$$\partial c/\partial t = \nabla \cdot [D_c \nabla c] + G\delta(\mathbf{r}) \qquad (3.4)$$

This equation is not difficult to solve. We get:

$$c(r,t) \sim \int_0^t \tau^{-1/2} e^{-r^2/4D\tau} d\tau \sim (\sqrt{t} - r\sqrt{\pi/4D}...) \qquad (3.5)$$

where we have written the limit for small *r*. That is, as in any diffusion problem, contours of constant *c* move according to *r~t^{1/2}*, *not with fixed velocity*. In order to get a fixed velocity we need a chemotaxis (drift) term like the second term in Eq.(3.1).

We also need an equation for the nutrient concentration, *n*. In the experiments that we have in mind there are several constituents of the nutrient used, but, for simplicity we will think about a single species such as glucose. Thus we have simple diffusion with a large diffusion constant, $D_n$, and consumption of the nutrient by the mobile cells, and also by those belonging to the MTS.

$$\partial n/\partial t = D_n \nabla^2 n - \kappa(n)c \qquad (3.6)$$

Here the last term corresponds to consumption of nutrient by the cells, and $\lambda(n)$, is a function giving the overall scale of nutrient consumption. Experiments with cultured brain cancer cells [21] show that $\lambda(n)$ is well represented by:

$$\lambda(n) = \begin{cases} \lambda n/n_1 & n \le n_1 \\ \lambda & n > n_1 \end{cases} \quad (3.7)$$

where $n_1$ is a characteristic saturation concentration. It is reasonable that $n_1 \ll n_o$, and we will make this assumption.

For the MTS, we assume the consumption is large, given the large number of highly proliferative cells in the surface layers. In order to avoid complications we set $n=0$ on the surface of the MTS. Far from the tumor we have a boundary condition: $n=n_\infty$, where $n_\infty$ is the nutrient concentration introduced at the beginning of the experiment, or, in the brain, the general supply.

Finally, we need an equation for the homotype factor, $h$. We use a diffusion equation, again, and suppose that this factor is produced by the mobile cells at a rate, $\nu$. In order to avoid a large buildup of this factor we assume that it decays at a rate, μ. Thus we have:

$$\partial h / \partial t = D_h \nabla^2 h - \mu h + \nu c \quad (3.8)$$

where $D_h$ is the corresponding diffusion coefficient. As we will see, in order to form branches, $D_h$ needs to be smaller than $D_c$.

## 3.2. Estimate of the Parameters

In order to proceed with a solution of the equations above, we need to know a number of parameters. In this section we will try to use experimental numbers insofar as possible and we will quote a number of quantities. These should be regarded as order-of-magnitude estimates for the specific experiment, which we are trying to explain, and still more so for the situation *in vivo* where many other effects such as inhomogeneity of the environment can play an important role. The point of the enterprise is to see whether the parameters we use in the simulation are reasonable.

We turn first to Eq. (3.6). The diffusion coefficient of glucose in the brain is known to be $D_n = 6.7 \times 10^{-7}$ cm$^2$/sec [22]. The saturation consumption has been measured, $\lambda =$ 1.6 pg/cell/min [21]. Also, $n_1$= 0.2 g/l. Note that this is within a factor of three of $n_\infty$ = 0.6 g/l for the experiment[1].

We should note that for our purposes Eq. (3.6) may simplified because the time scale for the diffusion of the nutrient is *much* faster than that of the cells, thus we may set $\partial n/\partial t = 0$, and write the equation in the following form (for $n > n_o$):

$$a^2 \nabla^2 (n/n_\infty) = [\Gamma_o a^2 / D_n n_\infty](\Phi c) \qquad (3.9)$$

Here we have measured $c$ in terms of the cell volume, $\Phi$, $n$ in terms of $n_\infty$, and multiplied by the square of a length scale (a typical cell diameter), $a$, which we take to be 10µm. The dimensionless group in brackets, $\Lambda_2 = \Gamma_o a^2 / D_n n_\infty$ turns out to be of order 0.1. It is a parameter that we need for our simulations.

For Eq. (3.1) we need $D_c$, the diffusion constant of brain tumor cells in tissue. The only direct measurement that we know of, by Burgess *et al.*[19], gives 1.7 x $10^{-9}$ cm²/sec. However, this value depends on interpreting data according to Eq. (3.2). We have explained above why we cannot accept this value from a tumor biology standpoint. Our point of view is that the velocity of the edge of the invasive zone, $v$, depends primarily on the chemotaxis, and is *independent of the random motion of the cells*.

We have been unable to find a direct measurement of the random diffusion of single invasive cells so that we are reduced to very crude estimates. We may state the problem as that of finding the 'jump time', $\tau$, i.e., the time required for a cell to move its own length in a random walk, $\tau = a^2/D_c$. For example, in the experiment of Chicoine *et al.*[3] a cell in a petri dish takes several hours to perform such a motion, and we may expect that in the brain, or within the gel medium of our experiment, the time will be longer. As a complete guess we suppose this time to be of order 10 hours. This amounts to guessing that $D_c \sim 10^{-12}$ cm²/sec. We are aware of the weakness of this chain of reasoning.

For the chemotaxis term we have a measurement[3] of the maximum drift velocity of mobile cells: $v$=4.8 µm/hr (which corresponds very well to the 4.5 µm/hr measured after 96 hrs in our 3D MTS assay[1]). Eq. (3.1) gives the drift velocity as:

$$v = \frac{1}{16} \chi n_\infty \nabla [n/n_\infty] \qquad (3.10)$$

Here the factor 1/16 comes from the receptor law. We may estimate the gradient in which the cells move by assuming that the nutrient recovers its full concentration in approximately the radius of the MTS, about 1mm. Thus $\Gamma_n \approx 2 \times 10^{-7}$ cm$^2$/sec. We note that this has the units of a diffusion constant, so we define our third dimensionless parameter, $\alpha_3 = \Gamma_n /D_c \approx 100$.

Much less is known about the dynamics of the homotype factor. We note that if a *steady state* is attained in Eq. (3.8), the ratio of the concentration of homotype factor, $h_o$, to cancer cells, $c_o$, is $h_o/c_o = \beta/\mu \equiv \alpha_4$, another dimensionless parameter. It will be convenient to measure the decay rate, $\mu$, in terms of the jump time, and set $\alpha_5 = D_c\mu/a^2 = \mu\tau$. We can parameterize the strength of the chemotaxis induced by $h$ by noting that $\alpha_6 = \chi h_o/D_c$ is a dimensionless group, which gives the relative importance of the first and last terms on the right-hand side of Eq. (1). Finally, we need to know the ratio of diffusion coefficients, $\alpha_7 = D_h/D_c$. These three parameters are not known, even in order of magnitude. The same is true of $\alpha_1$, which characterizes the homotype attraction in another way. In fact, understanding the role of these parameters, is the point of our enterprise here. We will attempt to deduce them from the pattern itself.

It is disturbing that we have seven parameters in this problem. However, we have been able to estimate two of them, and, as we will see, the general nature of the patterns is not terribly sensitive to the others. Also, we suggest that future experiments could be oriented to finding out more about the homotype factor.

## 4. Stability Analysis

In order to see whether chains of cells (forming the branches in Figure 1) are likely to be formed in our model, we start with a simplified two-dimensional continuum approach. Suppose that we consider a channel (a part of the invasive zone) with cells supplied at concentration $c_o$ at one end. Cells drift in the *x*-direction, with drift velocity, **u**, because of a fixed gradient of nutrient. The channel has a finite width, *L,* in the *y*-direction, and we use periodic boundary conditions. According to Eq. (3.8) there will be a steady state with a fixed concentration, $h_o = \alpha_4 c_o$, of homotype factor. The question we pose is whether this uniform steady state is stable. It turns out not to be, and the growth of the instability shows how branches start to form.

In our stability analysis we write:

$$c = c_o(1+C) \qquad h = h_o(1+H) \qquad (4.1)$$

where *C, H,* are small deviations from the steady state. We then rewrite linearized forms of Eqs. (3.1), (3.8) in terms of these variables. Further, we set $r = a$, using *a*, the cell diameter, to rescale spatial variables, and $t = (a^2/D_c)T$, scaling time by the jump time. We find:

$$\partial C/\partial T = \nabla^2 C - (a\mathbf{u}/D_c) \cdot \nabla C - \alpha_6 \nabla^2 H \tag{4.2}$$

$$\partial H/\partial T = \alpha_7 \nabla^2 H - \alpha_5(C - H) \tag{4.3}$$

In order to fix our ideas we will take a model set of parameters as follows: $\alpha_5 = .05$, corresponding to slow decay of the homotype agent, and $\alpha_6 = 5$, i.e., moderately strong chemotaxis from the homotype agent. The ratio of diffusion coefficients, $\alpha_7$, needs to be less than unity for a strong attraction to occur, so we take it to be 0.1. Finally, we can estimate the term $au/D_c$, from the previous section to be of order 3.

These are linear equations, and we seek a solution of the form:

$$C, H \sim \exp(\sigma T - i\mathbf{Q} \cdot \mathbf{r}) \tag{4.4}$$

Here the (dimensionless) growth rate of the instability, $\sigma$, plays the role of an eigenvalue, and the wavevector (in units of *1/a*), **Q,** controls the spatial variation. The dispersion relation, $\sigma(\mathbf{Q})$, gives the growth of instabilities for various wavelengths; positive $\sigma$ corresponds to unstable behavior. In Figure 2 we plot the dispersion relation for **Q** in the *x*-direction and the *y*-direction, for the parameters quoted above. The same general behavior is true over a large range in parameter space, namely instability at long wavelength (small **Q**), and stability at short wavelength.

We should note that the instability in the *y*-direction, i.e., across the channel, is stronger than that in the *x*-direction. In effect, the instability is advected by the drift velocity, **u**. In order to see the effect on an initial perturbation, we suppose that there is a small deviation from a uniform distribution of cells, and then propagate it forward in time using Eq. (4.2), (4.3). The equations can be solved exactly using Fourier Transforms, and this is what we did. The result is shown in Figure 3. The initial cell cluster, on the left, is carried forward by the drift, and increases in size by robbing material from the rest of the channel. This is incipient branch formation due to the combination of heterotype and homotype chemotaxis. Note that *both* are necessary to form this structure.

# 5. Discrete Simulation Model

## 5.1. Formulation

In the previous section we showed a linear analysis of the formation of a branch in the invasive zone. Here we will take a complementary approach and do a discrete simulation. In this case we represent the homotype factor by an extreme approximation: we assume it does not diffuse at all, so that each cell leaves a 'trail' which other cells follow. Alternately, we could interpret the model as representing the case where each cell carves a 'least resistance' pathway for others to follow.

We work on a 128x128 square lattice, with the lattice constant equal to *a*, the cell diameter. We treat the nutrient concentration as a continuous variable and the tumor cells as discrete. For the nutrient we solve Eq. (3.6) numerically on the lattice. If we write $N = n/n$, then Eq. (3.6) becomes:

$$\Delta_2 N = \alpha_2 C f(N) \tag{5.1}$$

where $f(N)=N$, $N <1/3$, $f=1$, otherwise (cf. Eq. (3.7)), $\Delta_2$ is a second numerical difference, and *C* is 1 or 0 depending on whether a cell occupies that lattice site. Eq. (5.1) is standard: it is a discrete version of the Helmholtz equation. We have solved it using the strongly implicit scheme.

For the cells, and represent the various processes by a set of jump probabilities. Our basic rate is the random walk rate for cells (c.f. the previous section). Each cell can jump to any empty adjacent site with rate $w(0) = 1$, unless it is either in a gradient of *N*, or if it undergoes homotype attraction. In the former case we put:

$$w(\nabla N) = \alpha_3 (3N +1)^{-2} \nabla N \tag{5.2}$$

if $\nabla N$, the numerical gradient of *N* in that direction, is positive. The second factor comes from the receptor law. For the homotype attraction, we keep track of the places where any cell has previously been. If that site has been visited we multiply each jump rate by $\alpha_1$. As the simulation proceeds we pick processes according to their relative jump rates. Showing that this discrete process reduces to Eq. (3.1) when viewed on large space and time scales is a standard exercise in stochastic analysis [23]. For the MTS we take a disk of immobile cells and start with about 100 mobile cells around it. In order to speed up the subsequent calculations we take a somewhat irregular initial distribution of these cells. In the course of the growth we allow any empty site near the immobile cells to either shed a new mobile cell, or to grow the MTS, with a certain relative probability. Our intention is to use this parameter to match the known growth history of the MTS, but we have not yet done this systematically.

## 5.2. Results

We show the results of simulation in three regimes in Figure 4 and Figure 5. In all cases we started with a tumor core of a radius of 20 cells, and liberated 100 cells to start invasion. We have taken $\mu_2$, the parameter which characterizes consumption of nutrient as 0.3, in the range we estimated above. The three images are representative of three regimes for the other two parameters. It will be important in future work to map out the various parameter regimes in more detail and see to what extent we can understand the underlying processes. The general scale of the figures can be thought of as being about 1mm, as in the experiment. In Figure 4a we have basically turned off both, chemotaxis and homotype attraction: $\mu_1 = 5$, $\mu_3 = 5$. In this case we have a typical result of diffusion: random walks are compact in two dimensions, and the cloud of cells near the MTS does not represent the invasive pattern of Figure 1. We conclude that it takes strong chemotaxis (or some equivalent phenomenon) to generate an invasive zone.

In Figure 4b we thus implemented strong chemotaxis, $\mu_3 = 100$, in the regime that we estimated above. We also took $\mu_1 = 25$, i.e., rather strong homotype attraction. In this pattern (Figure 4b) we see a dispersed zone of invading cells, yet only a hint of chain formation. However, this sort of pattern is observed in the experiment (cf. Figure 1b). It will be important to revisit the experiment in a systematic manner to try to understand how this pattern evolves.

We find that in order to produce well-defined chain structures, in Figure 5 (cf. Figure 1a), we need to introduce *very strong* homotype attraction, $\mu_1 = 250$, and also *very strong* heterotype chemotaxis, perhaps a factor of 10 larger than we estimated above. This might indicate that chemical triggers other than only glucose might be involved. With regards to the *in vitro* assay there are numerous candidate substances in the tissue culture medium as well as in the gel, and many more soluble factors are likely to have an impact in the real brain parenchyma. We are encouraged however by the fact that the model can produce chains at all without putting in an artificial device, which does not correspond to the biology that we understand.

## 6. SUMMARY

This preliminary study introduces a new computational model, which attempts to clarify the pattern of highly malignant tumor growth, in particular the origin and structure of the invasive zone. We have attempted to remain grounded in the biology of the tumor, yet we have, even in this preliminary study, already revealed a number of interesting features.

The nature of the pattern formation here is quite different from that assumed by other authors for related systems such as bacteria colonies[5]. In that case, the biology is different, in that bacteria reproduce while in motion. Thus the diffusive instability that gives rise to the branching shapes in bacteria colonies, which are very well represented by DLA. Our case is different despite the visual resemblance of the patterns, and the pattern formation is considerably more subtle. As such, this has model is interesting for statistical physics, quite apart from the application we have in mind. We should note, however, that our discrete simulation method is similar to that of Ben-Jacob *et al.* [5], and the diffusive instability is present here, as it is in the bacteria colonies: tumor cells that are far in advance of the rest of the tumor system consume nutrient, and tend to get even farther ahead.

Our results persuade us that chemotaxis is a driving force in forming the invasive zone, and this prediction can be tested. We propose that the invasive zone would slow its expansion considerably in situations where the main tumor mass is very well-nourished. The other ingredient in our model, the homotype attraction, is treated in a very schematic way here. In this form, it does give the qualitative effect that we are looking for, but the details of the process will need considerable work both in modeling and in understanding the underlying biological processes.

We are aware that several important features are not yet included in the model. For example, we have made no mention of haptotaxis. Haptotaxis refers to enhanced movement on a solid permissive substrate, e.g., the laminin and collagen fibers present within the matrigel used in here, and certainly *in vivo*. It is noteworthy that glioma tumor cells also produce extracellular matrix proteins such as laminin [24,25], thus further imprinting the pathway structure with increased permission on paracrine secreted, solid substrates. As such, aside from chemotaxis, haptotaxis also contributes to the observed branching patterns and thus needs to be considered for future work.

We have also completely neglected effects arising from the elasticity of the tissue surrounding the growing tumor. A growing core certainly strains the material, and this may affect the migrating cells, though the effect is not trivial since it can be shown that an expanding sphere in an elastic medium gives rise to a pure shear, i.e., local volumes are not changed, but rather deformed. We suspect that the important elastic effects are non-linear ones - this is quite likely in a gel, and presumably also in brain tissue. Aside from pure expansion, the mechanical effects of cell traction, i.e., tension, especially in the ECM-gel used in the experimental assay likely also need to be considered and our own preliminary experimental findings already support this notion. Another effect, which should be thought about in this context is the possibility of a kind of local fracture from the mobile cells enhancing the damage to the tissue, and favoring chain formation. In brain parenchyma, such micro-fractures may also lead to a decrease in the tissue consistency, thus reduce the mechanical confinement surrounding the main tumor mass. That would allow the tumor to continue its volumetric growth, which in turn would shed (in total) even more invasive cells. Interestingly, such a feedback pattern has indeed been observed *in vitro*, represented by the damped oscillations of the dynamic ratio between volumetric growth of the MTS and the invasive front[1].

In future work we intend to first map out in detail the parameter space for this *n* model, and to attempt to put in the observed growth profile of the MTS. We will then also study the effects of inhomogeneity of the 3D matrix, which should facilitate spatial expansion into directions of least resistance. More specifically, we plan to map out the response of the formation of the invasive zone to easier paths of motion (including haptotaxis). This effort will also help us determine more carefully the diffusion coefficient of brain tumor cells in such 3D heterogeneous biological media.

A better understanding of the processes governing the onset and the dynamics of multicellular tumor invasion would be of great significance for tumor biology research and an important step towards the development of novel diagnostic tools and innovative treatment approaches in the future.

**ACKNOWLEDGEMENTS:** LMS would like to thank Alexei Tkachenko and Gabriel Weinreich for helpful discussions. TSD would like to thank David A. Weitz, Daniel S. Fisher, and Michael E. Berens for fruitful discussions. Support from E. Antonio Chiocca for the Tumor Complexity Modeling Project is gratefully acknowledged. The human malignant glioma cell line U87MG EGFR was kindly supplied by Webster K. Cavenee (Ludwig Institute for Cancer Research, San Diego, CA).

## Figure Captions

**Figure 1.** Two microscopic images of human U87MG EGFR multicellular tumor spheroids (MTS) at t = 120 hours (after placement into the 3D ECM gel (= using the same gel composition in both experiments)). Figure 1a shows chain-like invasion patterns, whereas the MTS system in Figure 1b displays a more disc-like invasive pattern. Invasion in both MTS experiments is predominantly 2D because of the design of the Sandwich-Assay as described in detail in Deisboeck *et al.* [1].

**Figure 2.** Dispersion Relation for cells in a channel showing the stable and unstable modes for wavevector *Q*, in units 1/a, in the x and y directions. We show the growth rate (in the dimensionless units in the text) as a function of *Q*.

**Figure 3.** Growth instability within a channel. The initial cell cluster on the left grows into the incipient branch on the right. The axes are in units of *a*.

**Figure 4.** Simulation results. For Figure 4a, both chemotaxis and homotype attraction have been turned off. The resulting random walks are compact in two dimensions, however not representing the biological patterns. Figure 4b is the result of both strong chemotaxis and strong homotype attraction. The dispersed zone of invading cells with only a hint of chain formation resembles the experimental patterns observed in Figure 1b. The general scale of the figures can be thought of as being approximately 1 mm.

**Figure 5.** Simulation results. Very strong homotype attraction and very strong chemotaxis result in relatively well-defined chain-like structures as seen in Figure 1a. (The general scale can again be thought of as being approx. 1 mm).

**FIGURES**

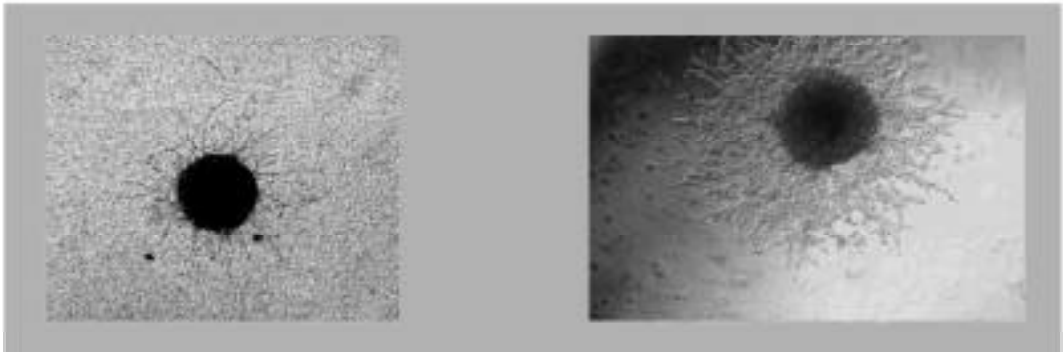

(a)      **(b)**

**Figure1**

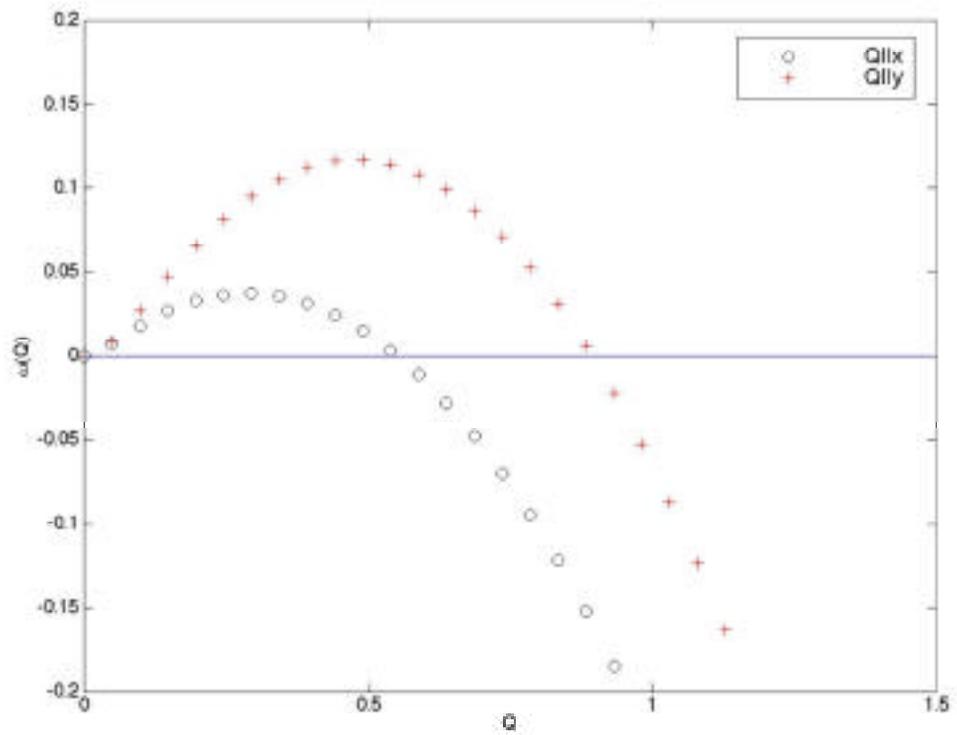

**Figure 2**

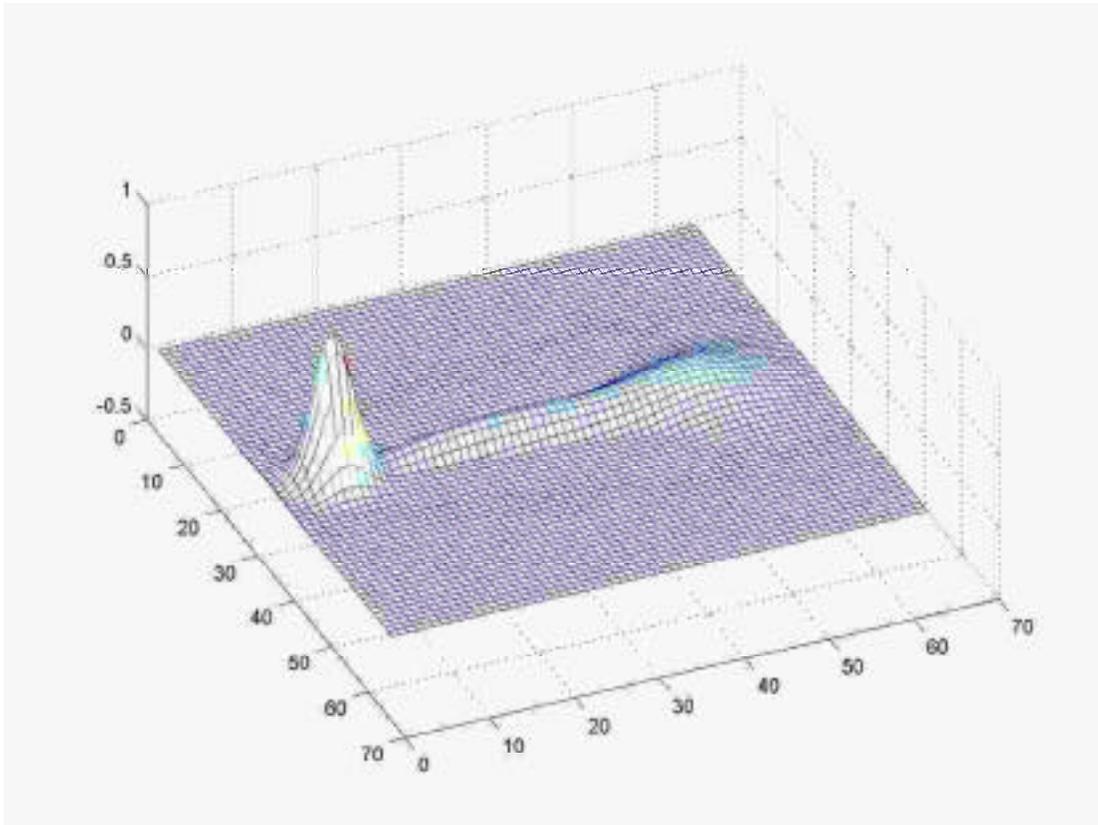

**Figure 3**

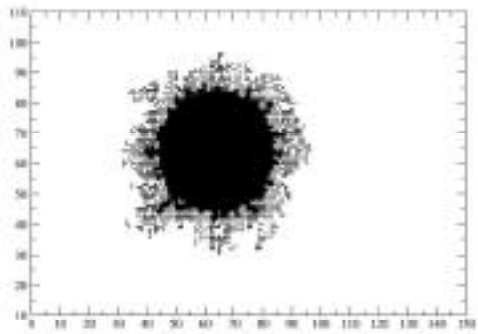 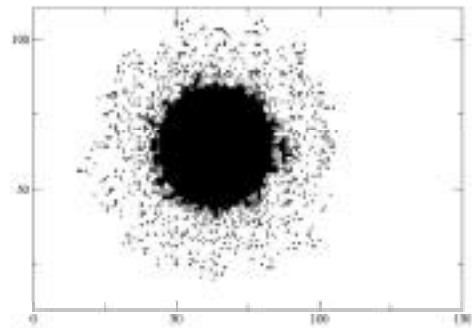

        **(a)**               **(b)**

**Figure 4**

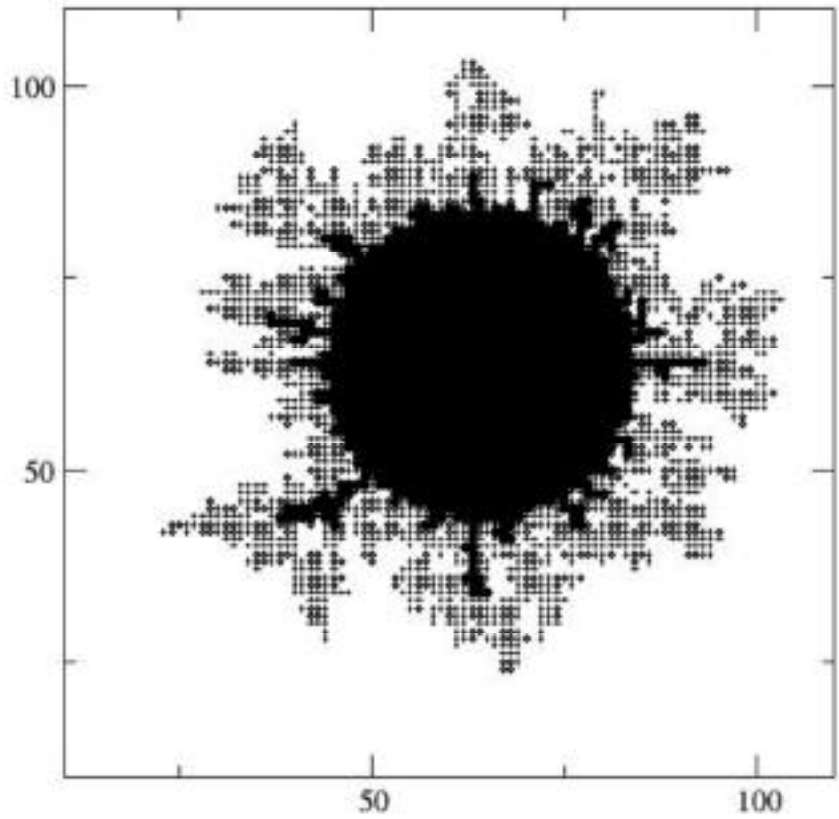

**Figure 5**